\documentstyle[11pt,paspconf]{article}

\input epsf

\begin{document}

\title{Cygnus X-1: A Case for a Magnetic Accretion Disk?}

\author{Michael A. Nowak\altaffilmark{1}, B. A. Vaughan\altaffilmark{2},
J. Dove\altaffilmark{3}, and J. Wilms\altaffilmark{4}}
\altaffiltext{1}{JILA, Campus Box 440, Boulder, CO ~80309-0440}
\altaffiltext{2}{Caltech, Pasadena, CA ~91125} 
\altaffiltext{3}{also Dept. of APAS, University of Colorado, Boulder, 
     CO ~80309}
\altaffiltext{4}{IAA University of T\"ubingen, Waldheuser Str. 64,
T\"ubingen, Germany}

\begin{abstract}
With the advent of RXTE, which is capable of broad spectral coverage and
fast timing, as well as other instruments which are increasingly being used
in multi-wavelength campaigns (via both space-based and ground-based
observations), we must demand more of our theoretical models.  No current
model mimics all facets of a system as complex as an x-ray binary.
However, a modern theory should qualitatively reproduce --- or at the very
least not fundamentally disagree with -- {\it all} of Cygnus X-1's most
basic average properties: energy spectrum (viewed within a broader
framework of black hole candidate spectral behavior), power spectrum (PSD),
and time delays and coherence between variability in different energy
bands. Below we discuss each of these basic properties in turn, and we
assess the health of one of the currently popular theories: Comptonization
of photons from a cold disk.  We find that the data pose substantial
challenges for this theory, as well as all other currently discussed
models.
\end{abstract}


\section{Energy Spectra}  

Historically, energy spectra have been the primary focus of both observers
and theorists.  The low state spectrum of Cyg X-1 (see Fig. 1) for the most
part is well-fit by a power-law with a photon index $\sim 1.7$.  Additional
facts to keep in mind are: the total luminosity of Cyg X-1 is likely $< 0.1
~ L_{\rm Edd}$, and that most other black holes show a strong soft
component and softer power-law tail above $> 0.1~ L_{Edd}$ (cf. Nowak
1995).  The softer, high state of Cyg X-1 is of roughly comparable
luminosity to the low state (Zhang 1996).  The soft components that do
exist in the low state are weak compared to the power-law.

The most popular models for this state are hot, optically thin models [with
advection dominated disks (ADD) being one example, cf. Narayan et
al. 1996], and Comptonization of seed photons from a cold disk (cf. Haardt
\& Maraschi 1993; Dove, Wilms, \& Begelman 1996).  ADD theories provide an
explanation for the hard to soft transition at $\sim 0.1 ~L_{Edd}$,
although they have yet to self-consistently calculate the (weak) observed
features that Comptonization models attribute to reflection.  One
theoretical motivation for the Comptonization models is the recent work 
(cf. Balbus, et al. 1996; Hawley et al. 1996) demonstrating the
existence of a powerful magnetic field-driven shear instability that is
strongest in low-density regions. Comptonization models assume that most of
the disk energy is dissipated in a corona, with roughly half being
reprocessed by the disk into soft photons (cf. Haardt \& Maraschi 1993).

Fitting the spectra of Cyg X-1 above 2 keV typcally requires $\tau_{es}
\sim 0.2$ and $kT_e \sim 150$ keV.  With roughly half of the luminosity
being reprocessed by the disk, and a large fraction escaping the optically
thin corona, we expect a large soft flux at $\sim 200$ eV.  This is {\it
not} observed (see Fig. 1 and Dove, Wilms, Begelman 1996, where more
refined models will be presented). In addition to not offering an explanation
for the low--high transition at $\sim 0.1~L_{Edd}$, this is a severe
problem for the corona models.  The models do, however, correctly account
for the Fe-line and reflection features, with $\sim$ half of the soft 200 eV
excess being due to reflection (as opposed to intrinsic disk emission).
ADD models, with a similar geometry and a strong source of hard photons,
may also overproduce soft X-rays if they attempt to model the observed
Fe-line with reflection off of the optically thick, geometrically thin
outer disk.

\begin{figure}

\centerline{\epsfxsize=0.5\hsize  {\epsfbox{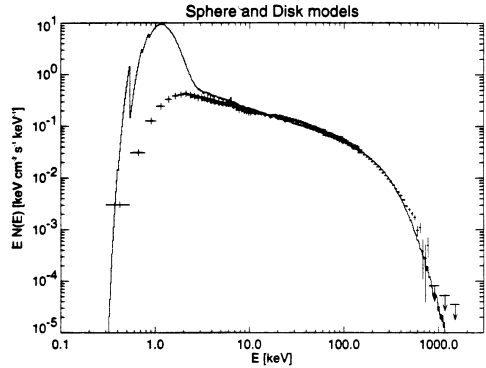}}}


\caption{Best-fit corona model to unfolded, non-simultaneous Cyg
X-1 spectrum.  Data are from BBXRT, TTM, (simultaneous with) HEXE, and OSSE
(Dove, Wilms, \& Begelman 1996).}  
\label{nowakm1.eps}
\end{figure}

\section{Phase/Time Lags Between Soft and Hard Variability}

Cyg X-1 shows strong X-ray variability (rms fluctuations $\sim 40\%$), with
timescales ranging from milliseconds to tens of seconds.  This variability
is usually measured by Fourier transforming the observed time series and
then forming the power spectral density (PSD), i.e. the squared transform.
One can also form the cross power density (CPD) by multiplying together the
complex transforms from two time series.  The Fourier frequency-dependent
phase of the CPD is known as the phase delay, $\theta_d(f)$, from which one
can form the time delay, $\tau_d \equiv \theta_d/2 \pi f$.  Cygnus X-1
shows a characteristic time lag of the hard photons behind the soft, with 
$\tau_d \sim 2 ~ (f/0.01 {\rm Hz})^{-2/3}~{\rm s}$ (cf. Miyamoto et al. 1992).
This corresponds to a nearly constant phase delay of $\sim 0.1$ rad. 

The only currently proposed ADD model explanation for this delay invokes
thermal disturbances propagating from the outer disk edge toward the center
(cf. Manmoto et al. 1996). This model requires, however, a very slow
propagation speed $\sim 10^{-4} c$ in order to be in quantitative
aggreement with observations (cf. Vaughan \& Nowak 1996).  Corona models
explain the delay as possibly arrising from the magnetic flares that are
allegedly energizing the corona (cf. Nowak 1994).  If the flare temperature
increases throughout its evolution, hard photons naturally lag soft
photons.  However, diffusion of photon scattering times within the corona
will lead to a ``smearing out" of this intrinsic time lag, depending upon
the spectrum of the seed photons, the corona temperature, optical depth, and
mean free scattering path (cf. Fig.2; Miller 1995; Nowak \& Vaughan 1996).
Current observations rule out large scattering paths ($\sim 10^8$ cm) with
very soft ($\sim 150$ eV) inputs.  Upcoming RXTE observations (obtaining
phase lags in 6-8 energy channels out to $\sim 250$ Hz) will provide more
stringent constraints.

\begin{figure}

\centerline{\epsfxsize=0.8\hsize  {\epsfbox{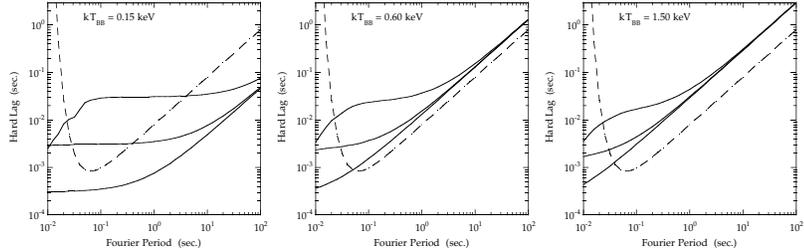}}}


\caption{Effects of Compton coronae on intrinsic phase lags, as a function
of input blackbody temperature and mean scattering path length
($\lambda_{es} = 10^6, 10^7,$ and $10^8$ cm).  From Nowak \& Vaughan
(1996). }
\label{nowakm2.eps}
\end{figure}

\section{Coherence Between Soft and Hard Variability}

The coherence function is essentially the normalized {\it average}
amplitude of the CPD (cf. Vaughan \& Nowak 1996, Nowak \& Vaughan 1996).
It is a measure of the fraction of the two time series that can be
interpreted as {\it time-independent and linear} transforms of one another.
Coherence must be less than or equal to unity, and be equal to unity only
when there is a linear, time-indpendent transfer function relating one
channel to the other.  Coherence is less than unity under any of the
following conditions: variability is due to thermal flares that are
dominated by emission from the Wien tale; there are multiple, independent
sources of variability (whether linear or not) with multiple,
independent responses to the sources (whether linear of not); there is {\it
any} non-linear transfer function from one channel to the other
(cf. Vaughan \& Nowak 1996).

If coronae are truly formed via multiple magnetic flares, then we expect a
strong loss of coherence, unless the the response to {\it each} of the
flares is identical.  Nowak (1994) showed that a kinematic model, with
multiple flaring regions, reproduced both the PSD and phase lags seen in
the very high state of GX339--4, as well as qualititatively agreed with the
energetics of coronal formation.  This model, and all others like it,
produces a strong loss of coherence (Fig. 3). This is {\it not} seen in Cyg
X-1 (Fig. 3; Vaughan \& Nowak 1996), which shows near unity coherence
between {\it all} well-observed energies.  This also disagrees with the
model of Manmoto et al. (1996).  For that case the disturbance shocks upon
reflection off of the disk inner edge, which is an inherently non-linear
process, and thus leads to less than unity coherence for a stochastic
source of variability.

\begin{figure}

\centerline{\epsfxsize=0.4\hsize  {\epsfbox{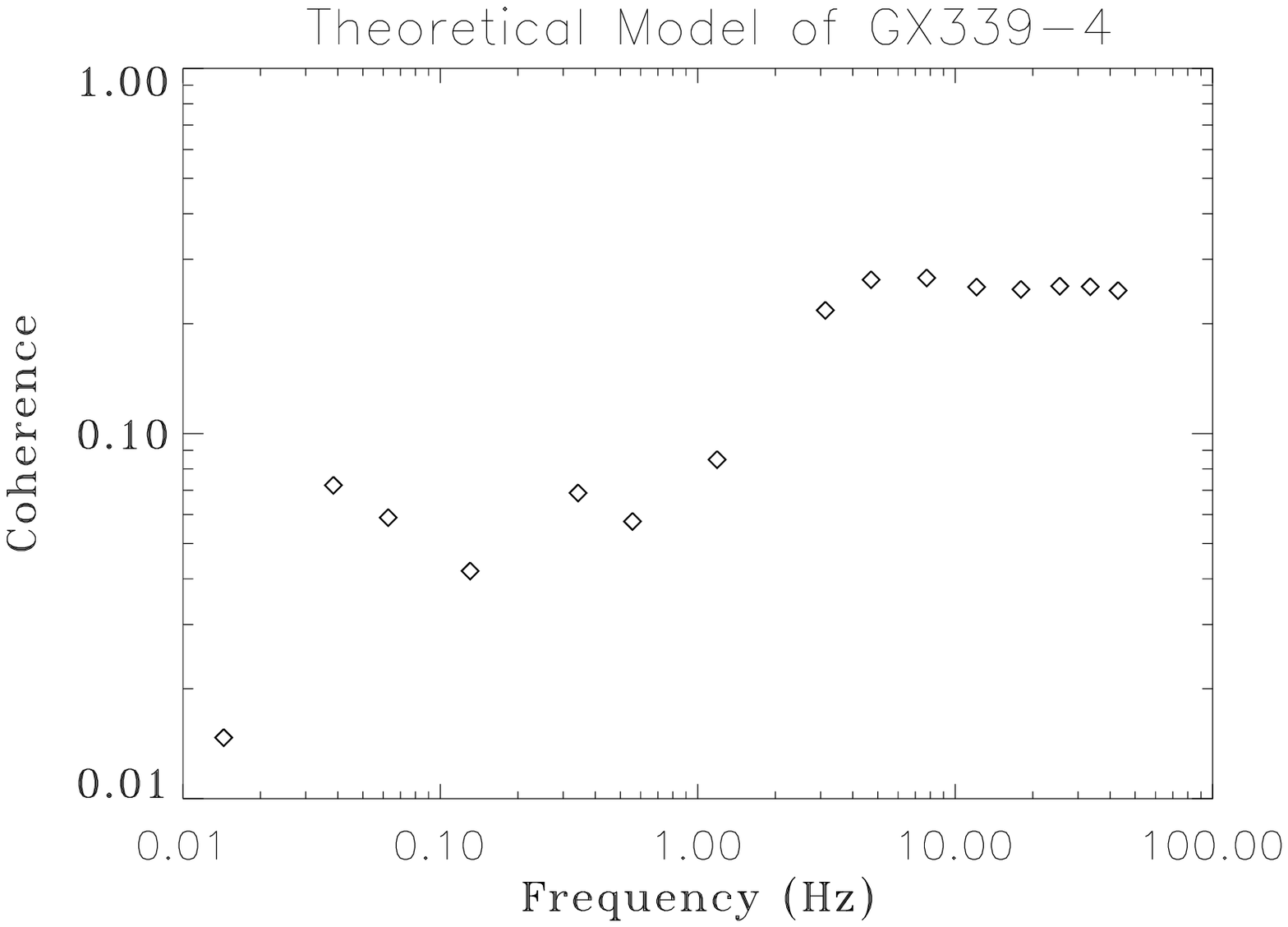}} ~~ 
\epsfxsize=0.4\hsize {\epsfbox{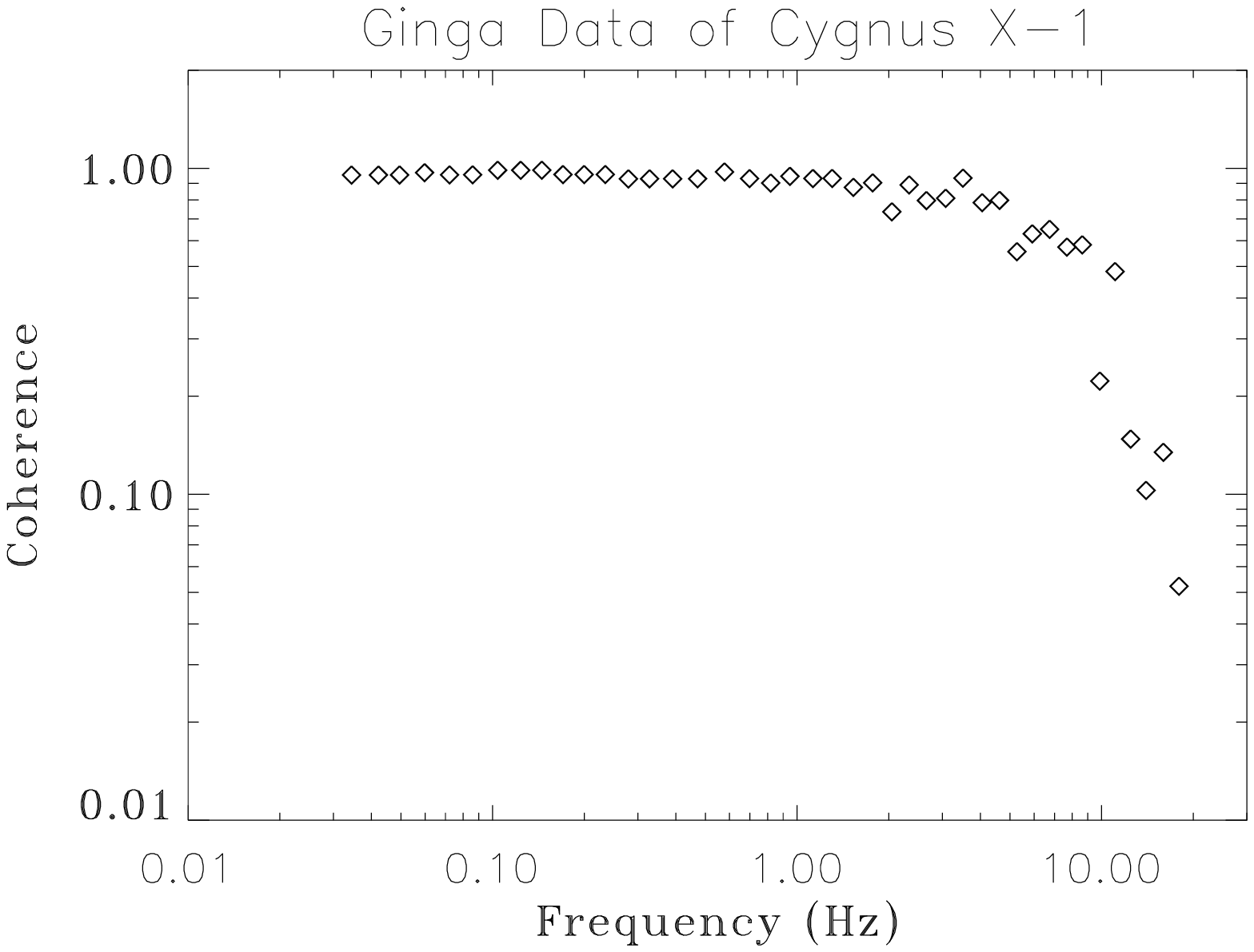}}  }


\caption{ Left: Coherence of theoretical model with multiple flares (based
on Nowak 1994).  Right: Ginga data of Cyg X-1 (cf. Vaughan \& Nowak 1996)
The Cyg X-1 loss of coherence is likely a noise artifact. }
\label{nowakm3a.eps,nowakm3b.eps}
\end{figure}

\bigskip

As they currently stand, coronal models of Cyg X-1 have several
problems. They predict too much soft X-ray flux, and they have yet to offer a
viable explanation for the phase lags that also preserves unity coherence.
ADD models also do not offer an explanation for the observed coherence, and
they have yet to self-consistently calculate the observed ``reflection" and
line features.

\acknowledgments

We would like to acknowledge useful conversations with M. Begelman,
P. Michelson, R. Staubert, C. Thompson, M. van der Klis, and all those who
weren't allowed to ask questions after the talk, but were good enough to
find me at coffee-break, especially F. Haardt and L. Maaraschi.
M.A.N. gratefully acknowledges support from NASA grant NAG 5-3225.


\begin{references}

\reference Balbus, S., et al. 1996, this volume
\reference Dove, J., Wilms, J., \& Begelman, M.C. 1996, \apj, submitted
\reference Haardt, F., \& Maraschi, L. 1993, \apj, 413, 507
\reference Hawley, J.C.., et al. 1996, this volume
\reference Manmoto, T., et al. 1996, \apjlett, 464, L135
\reference Miller, M.C. 1995, \apj, 441, 770
\reference Miyamoto, S., et al. 1992 \apjlett, 392, L21
\reference Narayan, R., et al. 1996, this volume
\reference Nowak, M.A. 1994, ApJ, 422, 688
\reference Nowak, M.A. 1995, \pasp, 718, 1207
\reference Nowak, M.A., \& Vaughan, B.A., 1996 \mnras, 280, 227
\reference Vaughan, B.A., \& Nowak, M.A., 1996 \apjlett, submitted
\reference Zhang, S.N., et al. 1996, this volume
\end{references}
\end{document}